\documentclass{PoS}

\usepackage{multicol}

\title{Twisted mass QCD at finite temperature}

\ShortTitle{Twisted mass QCD at finite temperature}

\author{Ernst-Michael~Ilgenfritz, Michael~M{\"u}ller-Preu{\ss}ker and Marcus~Petschlies\\
Humboldt-Universit\"at zu Berlin, Institut f\"ur Physik, Newtonstr.~15, 12489 Berlin, Germany \\
E-mail: \email{ilgenfri@physik.hu-berlin.de}, \email{mmp@physik.hu-berlin.de},
\email{marcuspe@physik.hu-berlin.de} }

\author{Karl~Jansen\\
DESY, Platanenallee 6, 15378 Zeuthen, Germany \\
E-mail: \email{Karl.Jansen@desy.de}}

\author{Maria Paola~Lombardo\\
Istituto Nazionale di Fisica Nucleare, LNF, Via Enrico Fermi 40, I 00044, Frascati (Roma), Italy\\
E-mail: \email{Lombardo@lnf.infn.it}}

\author{Owe~Philipsen and \speaker{Lars~Zeidlewicz} \\
Westf\"alische Wilhelms-Universit\"at M\"unster, Institut f\"ur Theoretische Physik,
Wilhelm-Klemm-Str.~9, 48149 M\"unster, Germany \\
E-mail: \email{ophil@uni-muenster.de}, \email{zeidlewicz@uni-muenster.de}}

\author{Andr\'e~Sternbeck\\
CSSM, School of Chemistry \& Physics, The University of Adelaide,
SA 5005, Australia \\
E-mail: \email{andre.sternbeck@adelaide.edu.au}}

\abstract{
We discuss the use of Wilson fermions with twisted mass for 
simulations of QCD thermodynamics. As a prerequisite for a future analysis 
of the finite-temperature transition making use of automatic $\mathcal{O}(a)$ 
improvement, we investigate the phase structure in the space spanned 
by the hopping parameter $\kappa$, the coupling $\beta$, and the 
twisted mass parameter $\mu$. We present results for $N_f=2$ degenerate 
quarks on a $16^3 \times 8$ lattice, for which we investigate the 
possibility of an Aoki phase existing at strong coupling and vanishing $\mu$, 
as well as of a thermal phase transition at moderate gauge couplings and 
non-vanishing $\mu$.
}

\FullConference{The XXV International Symposium on Lattice Field Theory\\
July 30 - August 4, 2007\\
Regensburg, Germany}

\begin{document}
\section{Introduction}

Wilson fermions with a twisted mass term evaluated at ``maximal twist''
offer a number of advantages over untwisted fermions, such as the
absence of exceptional configurations and automatic $\mathcal{O}(a)$
improvement. In order to see whether these features can also be
successfully exploited for future analyses of the QCD finite
temperature phase transition, we present as a prerequisite an
investigation of the phase diagram spanned by the hopping parameter
$\kappa$, the twisted mass parameter $\mu$ and the gauge coupling
$\beta$ for $N_f=2$ quarks.

The twisted mass action for two degenerate fermions consists of the
standard Wilson action, augmented by a twisted mass term
$\overline{\psi}i\mu\gamma_5\tau^3\psi$.  The hopping parameter
$\kappa$ and the twist parameter $\mu$ are related to the bare quark
mass as
\begin{equation} m_q = \sqrt{\frac{1}{4}\left(
\frac{1}{\kappa}-\frac{1}{\kappa_c}\right)^2 + \mu^2 }.
\label{eq:mass}
\end{equation} For maximal twist we have $\kappa=\kappa_c$, in which
case the quark mass is determined by $\mu$ alone. For a recent review
of the twisted mass formulation, see \cite{Shindler:2007vp}.

%\subsection{Expected Phase Structure}

\begin{figure}[h]
\centering
\begin{minipage}[t]{0.45\textwidth}
\centering
\includegraphics[width=0.8\textwidth]{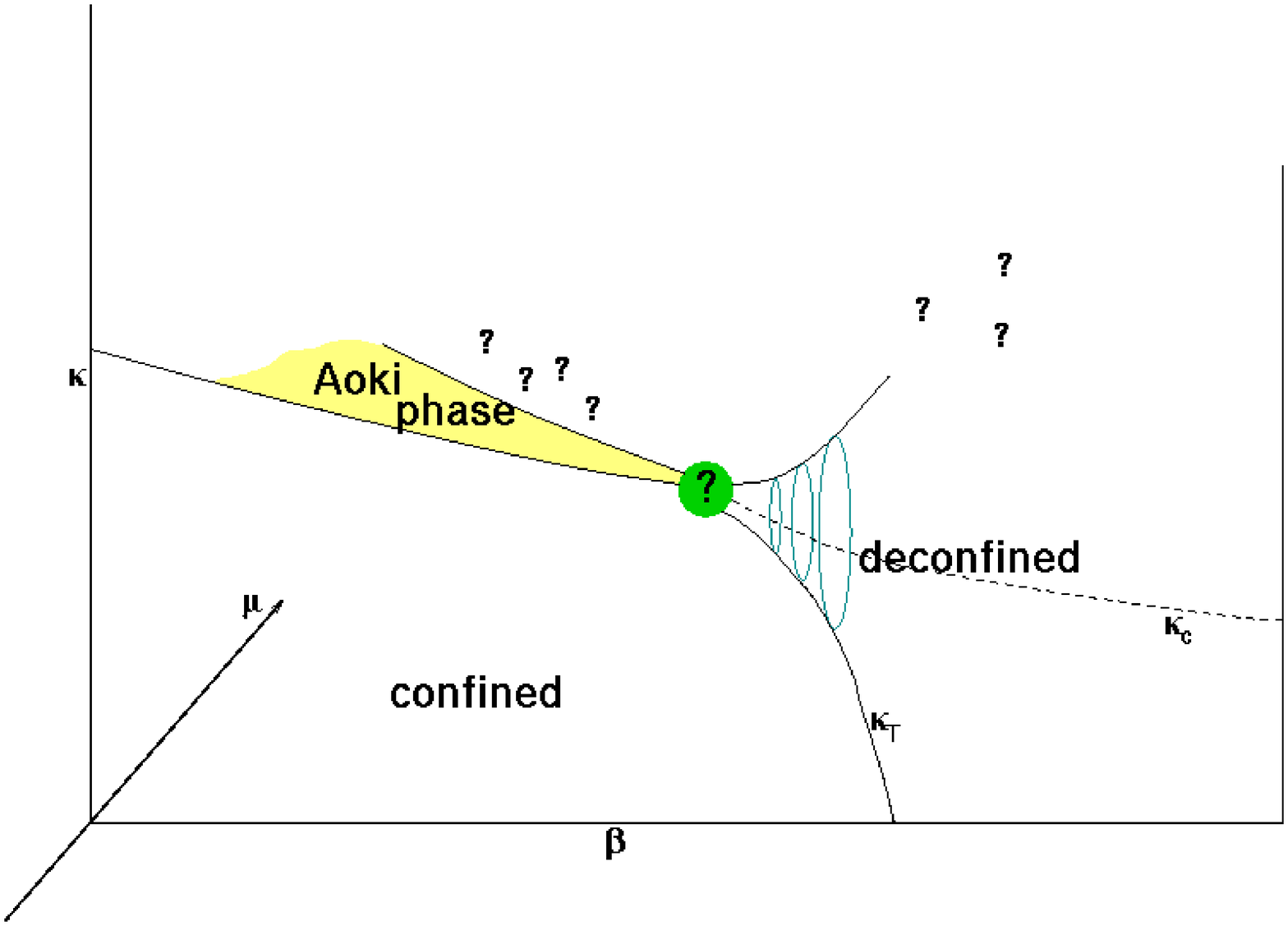}
\caption{Speculative phase structure for twisted mass Wilson fermions. 
The Aoki phase is part of the $\mu=0$ plane and the thermal transition
is expected to be located on a conical surface;
cf.~\cite{Creutz:2007fe}.\label{creutzphase}} 
\end{minipage}
\hspace*{2mm}
\begin{minipage}[t]{0.45\textwidth}
\centering
\includegraphics[width=0.8\textwidth]{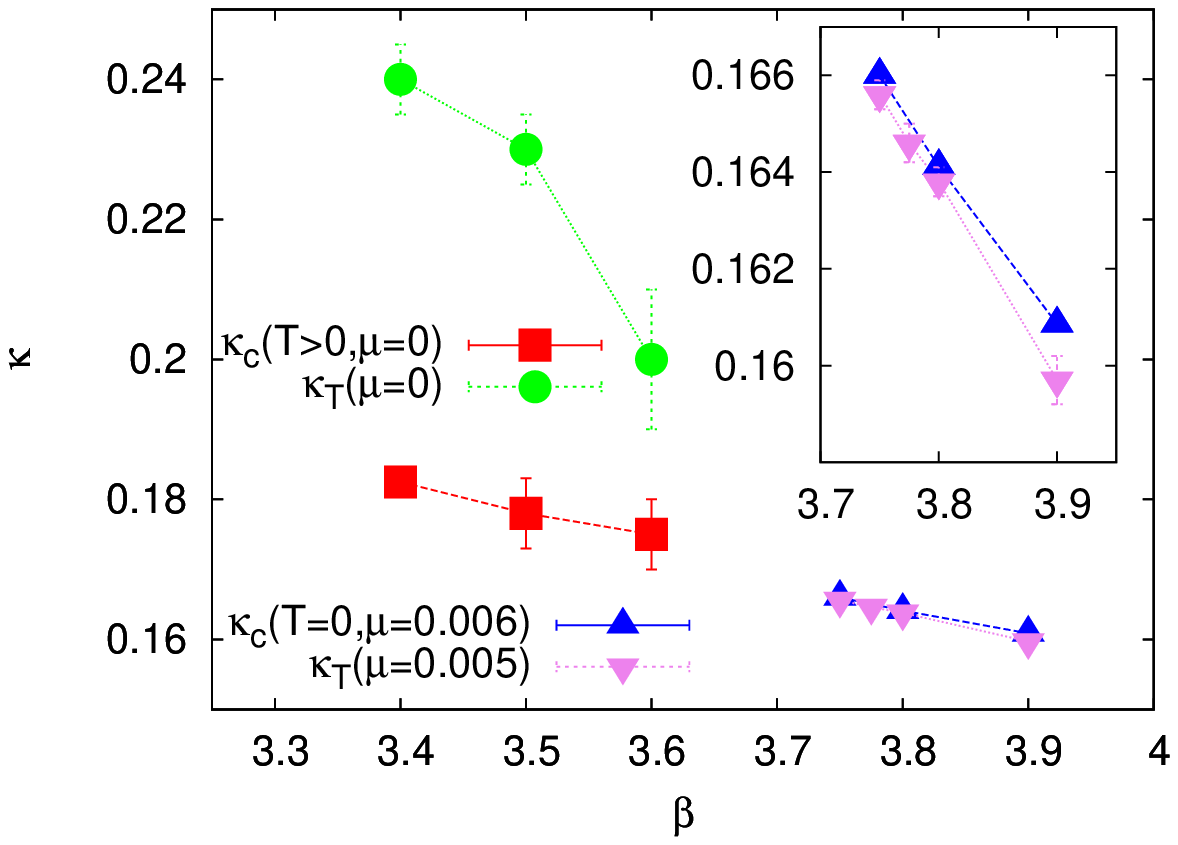}
\caption{Summary of results. Note that the left part of the diagram shows the $\mu=0$ plane whereas the right part (also the enlarged insert) shows the thermal transition line for $\mu=0.005$.\label{summplot}}
\end{minipage}
\vspace*{-.1cm}
\end{figure}

The expected phase structure is based on symmetry arguments and
leading order chiral perturbation theory \cite{Sharpe:1998xm,Munster:2004am,Farchioni:2004us}. It has
recently been discussed in \cite{Creutz:2007fe}, a qualitative sketch
is shown in Figure~\ref{creutzphase}.  The phase diagram divides into
two regions. The expected Aoki phase for small values of $\beta$
(cf.~\cite{Ilgenfritz:2005ba}) is in the $\mu=0$ plane, since $\mu\neq
0$ explicitly breaks the associated parity-flavour symmetry.  For
larger values of $\beta$, a conical surface is believed to mark the
thermal transition between the hadronic and quark gluon regimes.  This
is based on the expectation \cite{Creutz:2007fe} that, for fixed
$\beta$, the lines of equal bare quark mass in the $(\kappa,\mu)$
plane given by equation~(\ref{eq:mass}) should correspond to lines of
constant physics at that lattice spacing.

Our simulations are performed on a $16^3\times8$ lattice using an
improved HMC code as described in \cite{Urbach:2005ji}, based on the
tree-level Symanzik improved gauge action.  We search for signatures
for the Aoki phase and its possible decoupling from the deconfinement
transition in the range from $\beta=3.40$ to $3.60$.  For larger
$\beta$, we present results for the slice of the surface at fixed
$\mu=0.005$ obtained from simulations at $\beta\in\{3.75,3.775,3.8\}$,
supplemented by data at $\beta=3.9$ from a previous investigation
\cite{Ilgenfritz:2006tz}. Our findings, to be discussed in the
following sections, are summarized in Figure~\ref{summplot}.

\section{Search for the Aoki Phase}
\begin{figure}[p]
\centering
\begin{minipage}[t]{.45\textwidth}
\includegraphics[width=.9\textwidth]{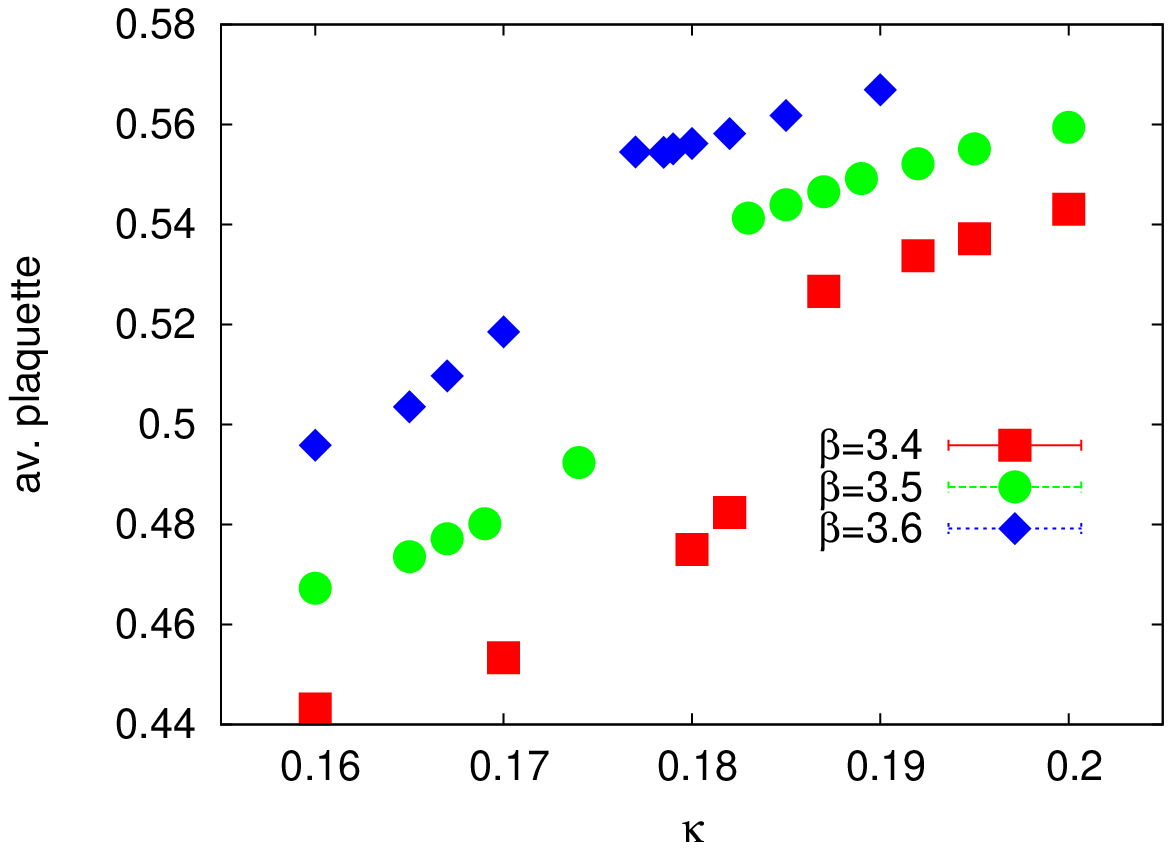}\\
\includegraphics[width=.9\textwidth]{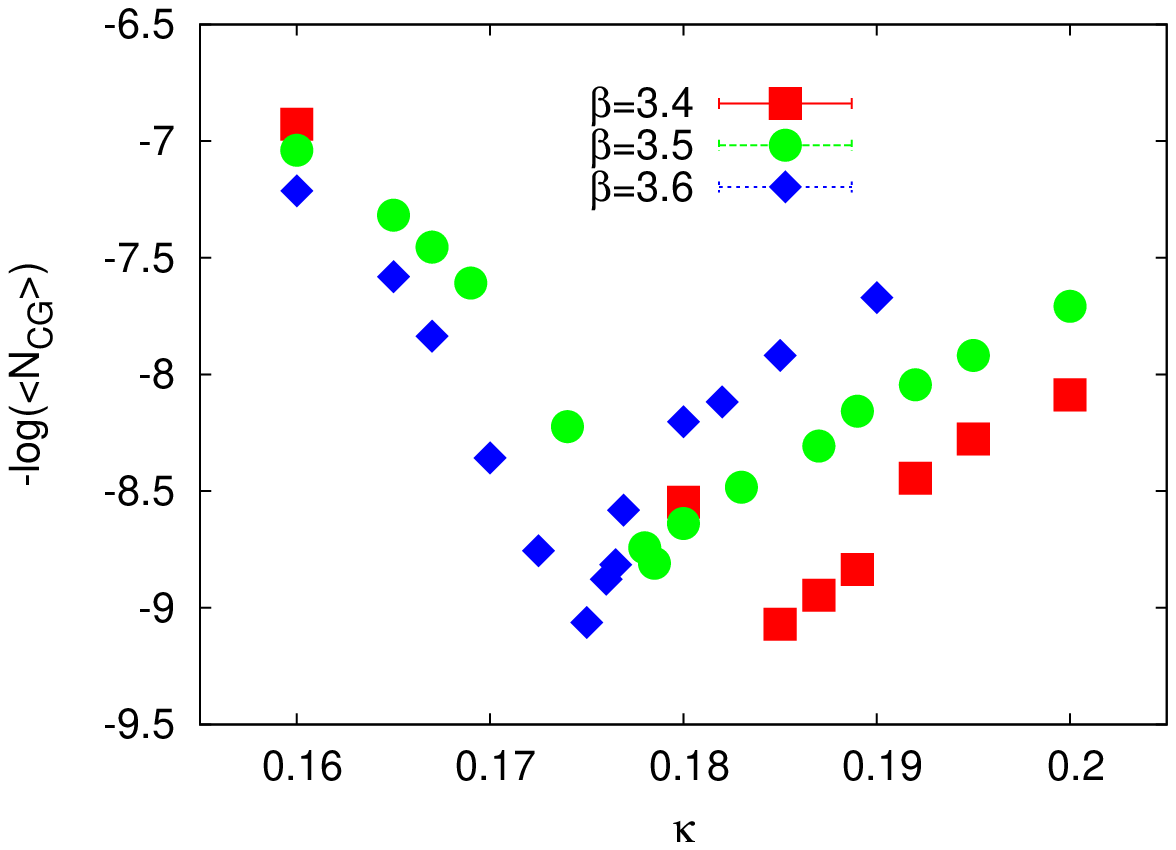}
\end{minipage}
\begin{minipage}[t]{.45\textwidth}
\includegraphics[width=.9\textwidth]{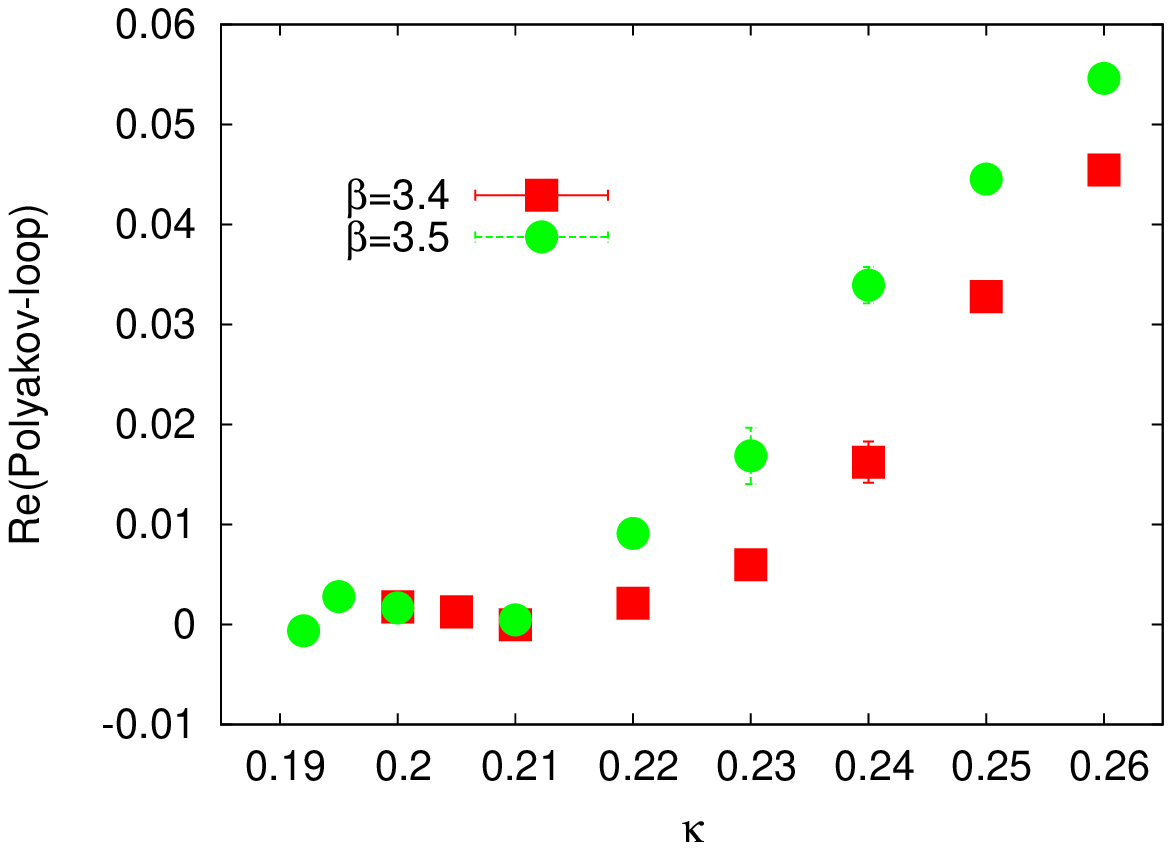} \\
\includegraphics[width=.9\textwidth]{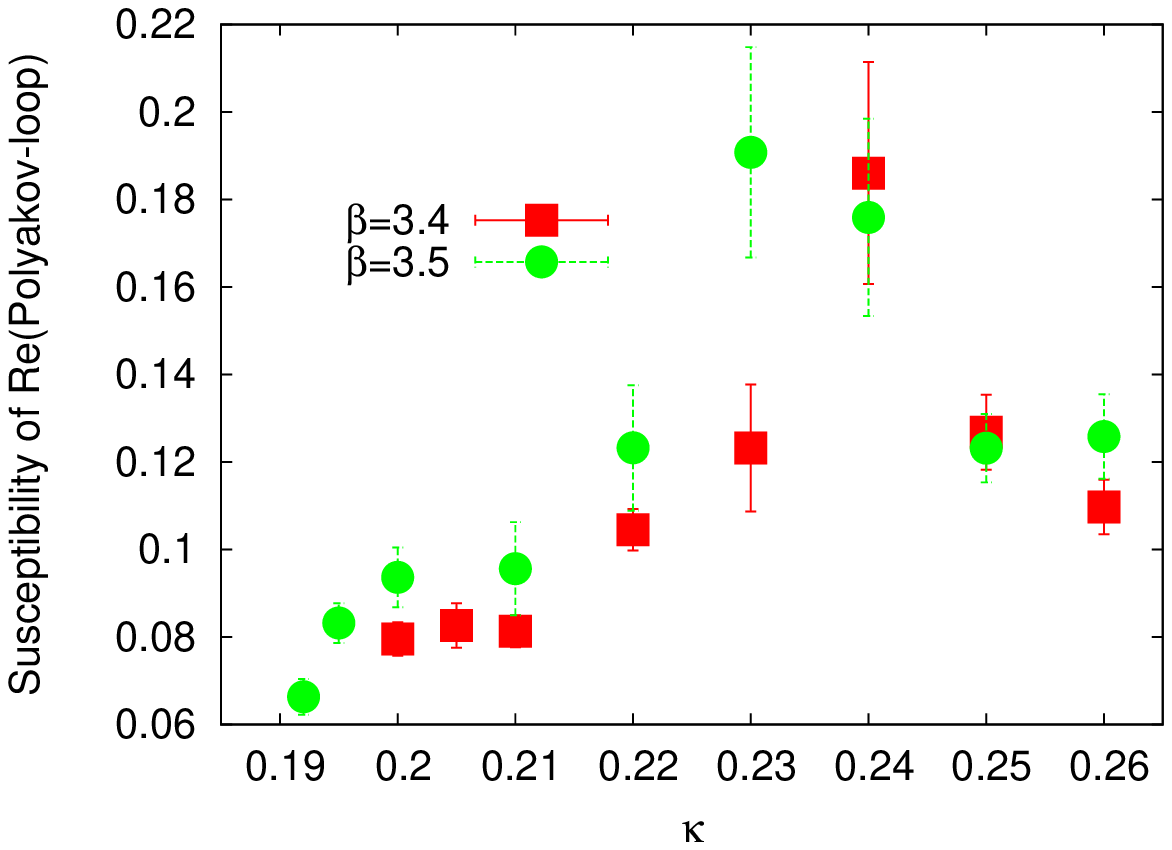}
\end{minipage}
\caption{Scan of the phase diagram at $\mu=0$ looking for signals of an
Aoki phase.  \textbf{Left:} Average plaquette (top) and
minus the logarithm of the average number of CG-iterations (bottom) as
functions of $\kappa$. \textbf{Right:} The real part of the Polyakov
loop (top) and its susceptibility (bottom) also as functions of $\kappa$. 
\label{mu0figs}}
\vspace*{0.8cm}
\begin{minipage}[t]{.44\textwidth}
\includegraphics[width=.9\textwidth]{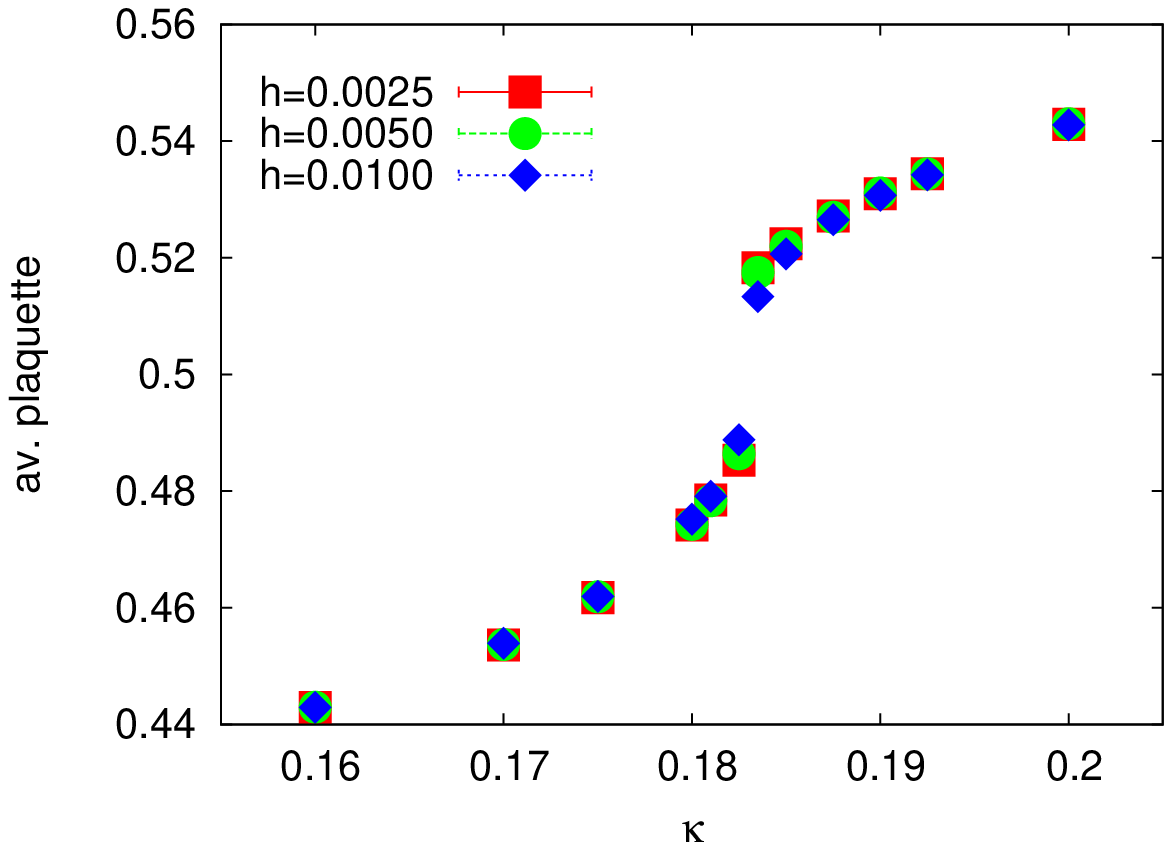}\\
\includegraphics[width=.9\textwidth]{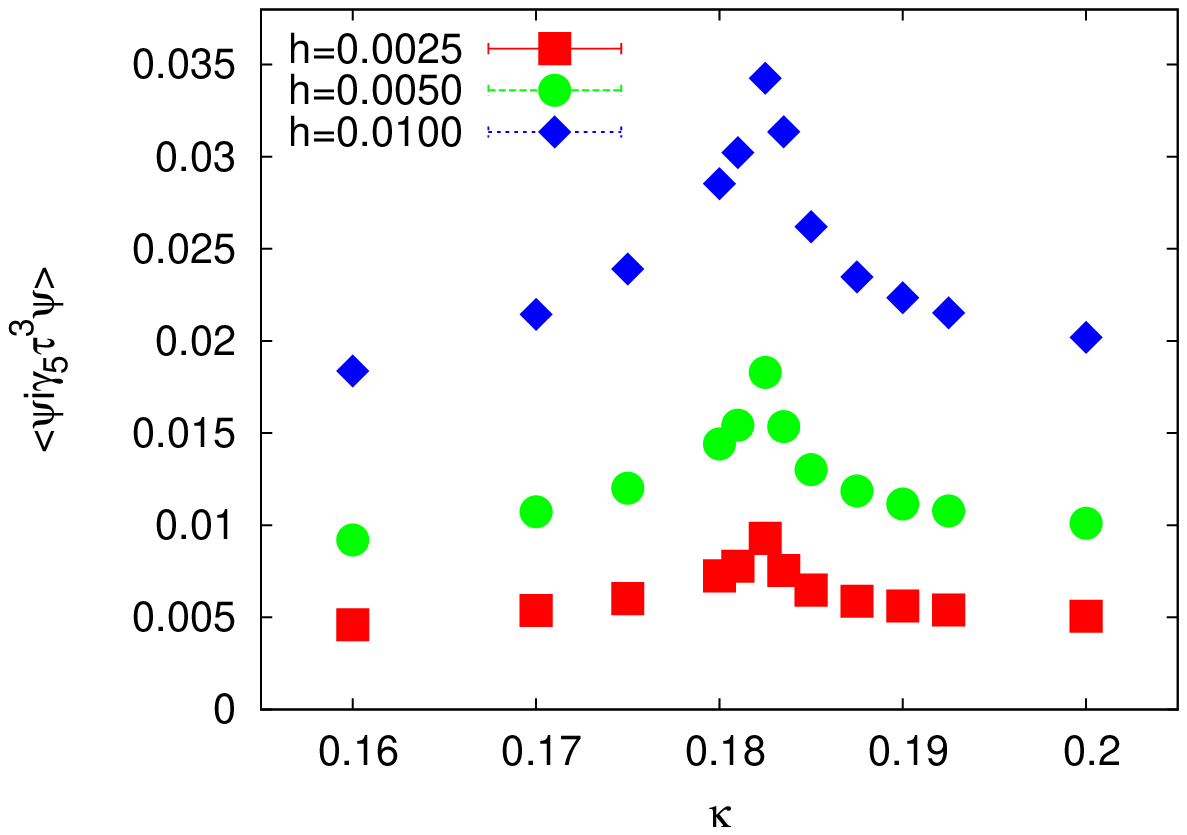}
\end{minipage}
\begin{minipage}[t]{.44\textwidth}
\includegraphics[width=.9\textwidth]{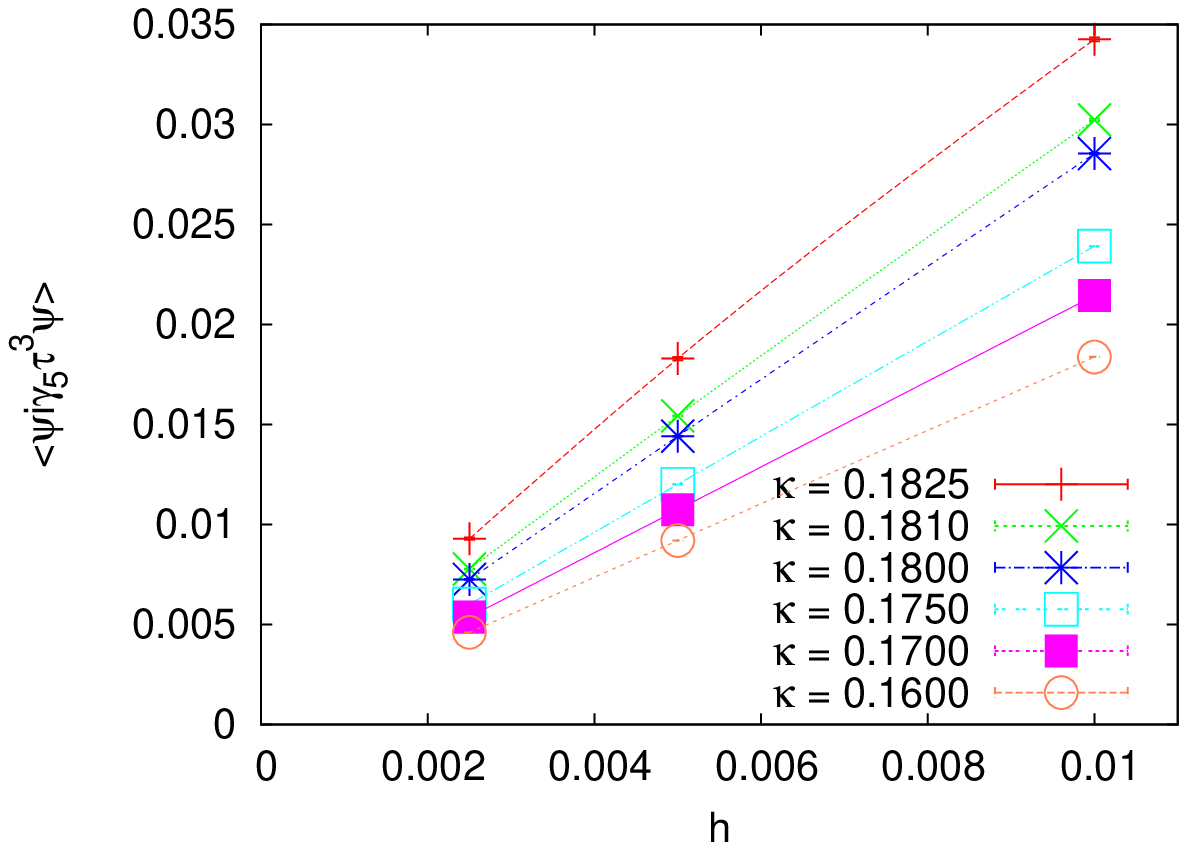} \\
\includegraphics[width=.9\textwidth]{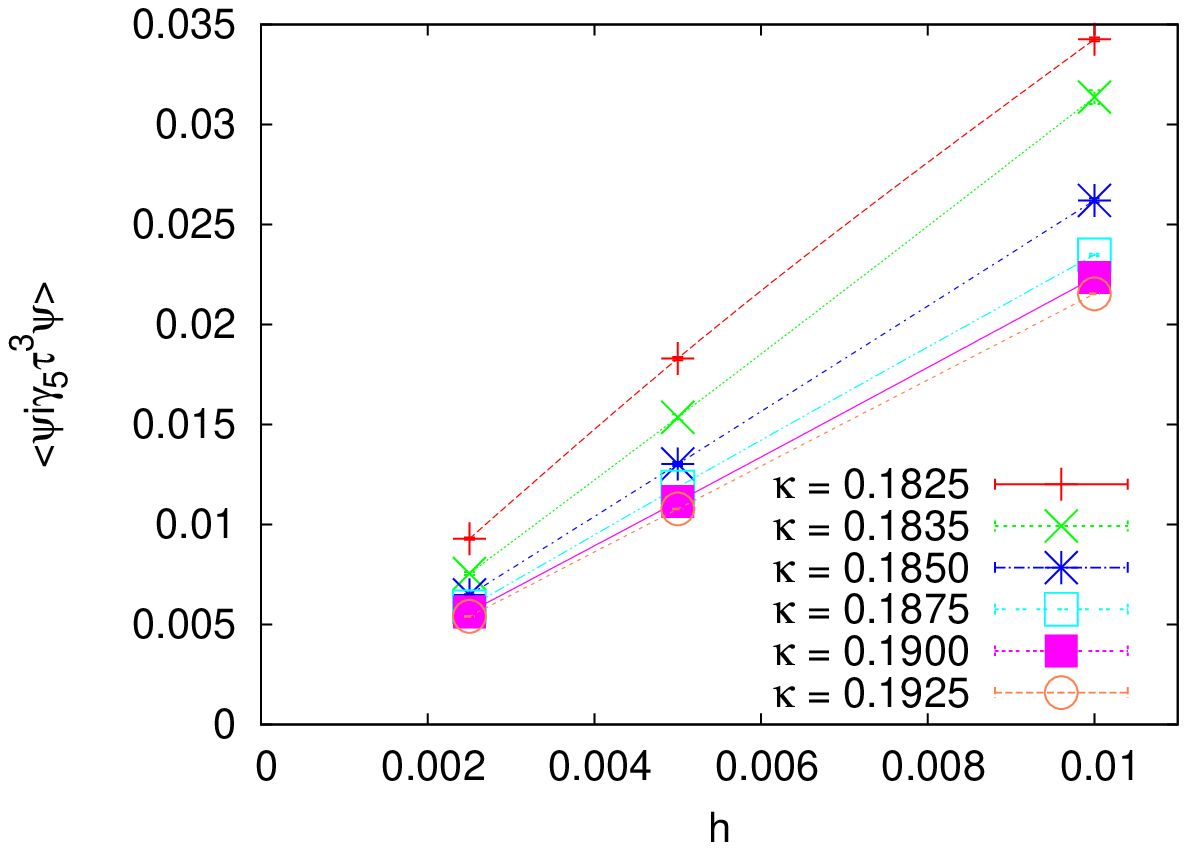}
\end{minipage}
\caption{Search for the Aoki-phase at $\beta=3.4$ at 3 different
values of $h$.  \textbf{Left:} Average plaquette (top) and the order
parameter (bottom) versus $\kappa$.  \textbf{Right:} The order
parameter versus $h$ for $\kappa\le\kappa^*$ (top) and
$\kappa\ge\kappa^*$ (bottom).
\label{hgt0figs}}
\end{figure}

In a first attempt to roughly estimate a region of parameter space for a potential Aoki phase, we performed a scan in $\kappa$ at several values of $\beta$ keeping $\mu=0$ fixed. According to the Aoki phase scenario the masses of the charged as well as the neutral pions vanish along the border of the Aoki phase. Consequently, we expect the average number of conjugate-gradient iterations $\langle N_{CG} \rangle$ in the HMC process to peak there. Such a behaviour is demonstrated in the lower-left graph of Figure~\ref{mu0figs} where
$-\log \langle N_{CG} \rangle$ is shown as a function of $\kappa$ for different values of $\beta$. We find that the minimum of $-\log \langle N_{CG} \rangle$ shifts slightly towards lower $\kappa$ as $\beta$ increases. In the upper-left panel of Figure~\ref{mu0figs}, this is consistently reflected in the behaviour of the average plaquette, showing a rather rapid change when approaching the region where $\langle N_{CG}\rangle$ peaks. Therefore, given the data of $\langle N_{CG}\rangle$, a possible Aoki phase should lie within the narrow interval $\kappa\, \in\, [0.16,0.20]$.

The Polyakov loop as an indicator for a thermal phase transition remains flat throughout this interval. A significant rise can only be found for $\kappa \ge 0.21$, i.\,e. well beyond the possible Aoki phase region. This can be seen on the right hand side of Figure~\ref{mu0figs} where the real part of the Polyakov loop and its susceptibility are shown as functions of $\kappa$. Therefore, if an
Aoki phase were found somewhere between $\kappa=0.16$ and 0.20, this would be in contrast to the experience gained in previous investigations (with Wilson gauge action) \cite{Ilgenfritz:2005ba} in which no clear separation of the Aoki phase from the thermal transition has been observed. As
expected for the phase diagram at $\mu=0$, the $\kappa$ value at which the Polyakov loop rises significantly shifts to lower values when
$\beta$ is increased.

At $\beta=3.4$ we have started a more thorough investigation of the existence of a phase of broken parity-flavour symmetry by searching
for a $\kappa$-region where the order parameter $\langle\bar{\psi}i\gamma_5\tau^3\psi\rangle$ does not vanish in the limit
$\lim_{h\to0}\lim_{V\to\infty}\langle\bar{\psi}i\gamma_5\tau^3\psi\rangle$. Here the strength of the ``external field'' $h$ is related to the twisted mass parameter $\mu$ by $h = 2\kappa\mu$. So far, we have only measured the order parameter as a function of $\kappa$ at $\beta=3.4$ and the three values $h=0.01$, $0.005$ and $0.0025$. Our current results for the average plaquette and the order parameter are shown in Figure~\ref{hgt0figs}; on the left hand side both observables as functions of~$\kappa$ and on the right hand side only $\langle\bar{\psi}i\gamma_5\tau^3\psi\rangle$ versus~$h$. As above, the average plaquette exposes a rapid change or discontinuity as a function of~$\kappa$. This becomes slightly stronger with decreasing $h$. The order parameter is observed to peak around $\kappa^*\approx 0.1825$. This peak becomes smaller and narrower in width as $h$ decreases. Looking at the right hand side of Figure~\ref{hgt0figs}, we get strong indications that for $h \to 0$ the order parameter tends to zero for all $\kappa$ values considered. Therefore, our yet preliminary results at $\beta=3.4$ point to an absence of the Aoki phase and a possible first order phase transition, a scenario sustained by the occurence of long-lasting metastable states at $\kappa=0.1830$. In this context it should be mentioned that first order behaviour for Wilson fermions in the intermediate coupling regime has been observed before (cf.~\cite{Iwasaki:1994gq} and references therein). We interprete our data in the same way as a remnant of the $\kappa_c(T=0)$ bulk transition. Of course, the existence of an Aoki phase at smaller $\beta$ is left open. In any case, our results for the Polyakov loop show that the deconfining transition happens at considerably larger $\kappa$ values.

\newpage
\section{Thermal Transition Line}

In order to locate the thermal transition line for $\mu=0.005$ we
fixed $\beta$ and scanned in $\kappa$, reaching $\mathcal{O}(10000)$
statistics for each pairing.  As observables we used the plaquette,
the real part of the Polyakov loop and their respective
susceptibilities as well as the pion norm, which is given by
$\Vert\pi\Vert^2 = \sum_x\textrm{Tr} \,
\overline{\psi}(x)\gamma_5\psi(x)\overline{\psi}(0) \gamma_5\psi(0)$.
The susceptibilities have been normalized as $\chi_O = N_s^3\left(
\left<O^2\right>-\left<O\right>^2\right)$.  The critical coupling
$\kappa_T$ for the thermal transition is then signalled by a peak in
the susceptibilities.

\begin{figure}[h]
\centering
\begin{minipage}[t]{.45\textwidth}
\centering
\includegraphics[width=.9\textwidth]{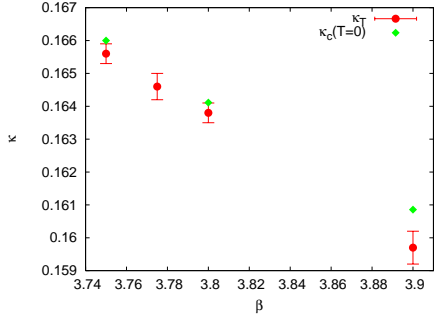}
\end{minipage}
\begin{minipage}[t]{.45\textwidth}
\centering
\vspace*{-4.4cm}
\caption{Results for the thermal transition line $\kappa_T(\beta,\mu=0.005)$ compared 
to the known points for the critical $\kappa_c$ at $T=0$.\label{thermalline}}

\small\vspace*{0.2cm}
\begin{tabular}{c|c|c}
$\beta$ & $\kappa_T(\beta,\mu=0.005)$ & $\kappa_c(T=0,\beta)$   \\
\hline
3.75  &  0.1656(3)  &  0.1660(1)  \\
3.775 &  0.1646(4)  &  --         \\
3.8   &  0.1638(3)  &  0.164111   \\
3.9   &  0.1597(5)  &  0.160856   
\end{tabular}
\end{minipage}
\end{figure}

\begin{figure}%[p]
\centering
\begin{minipage}[t]{.45\textwidth}
\includegraphics[width=.9\textwidth]{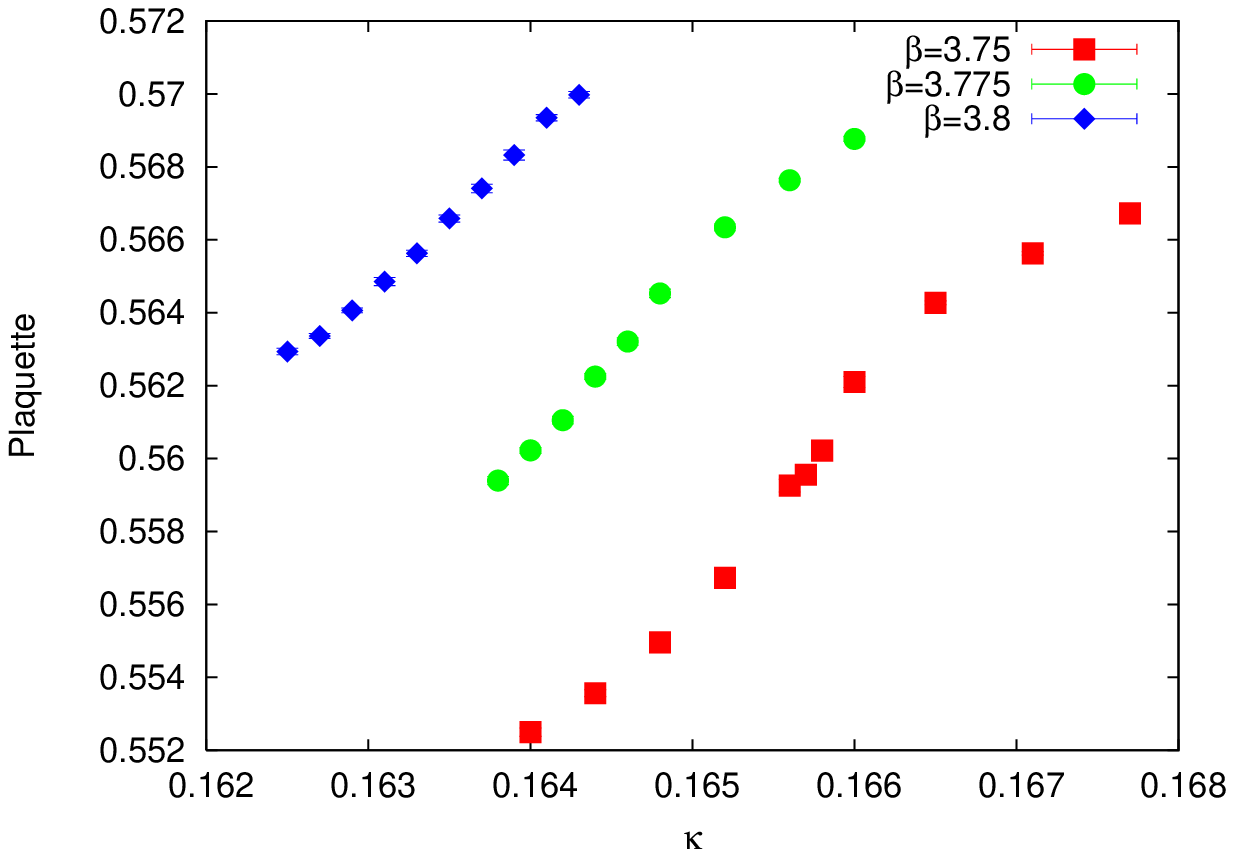} \\
\includegraphics[width=.9\textwidth]{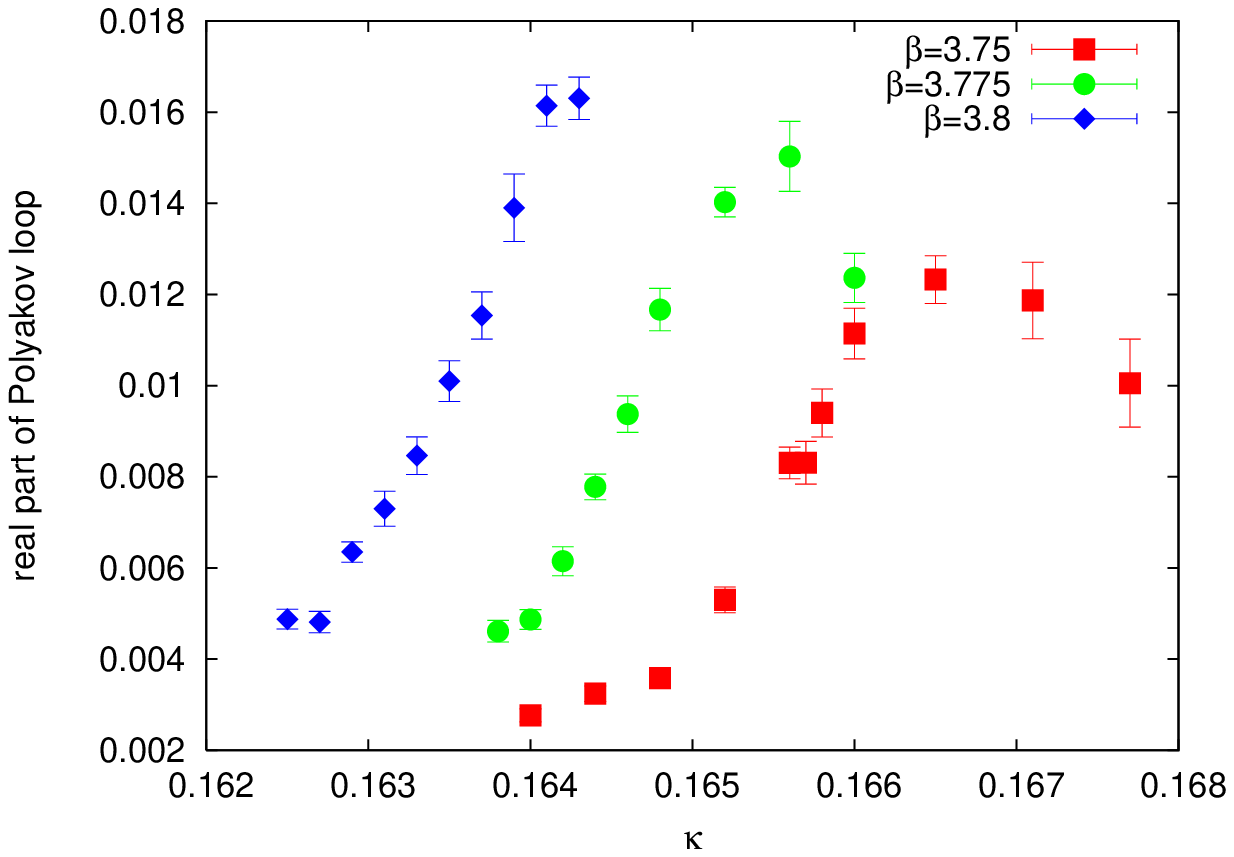} \\
\includegraphics[width=.9\textwidth]{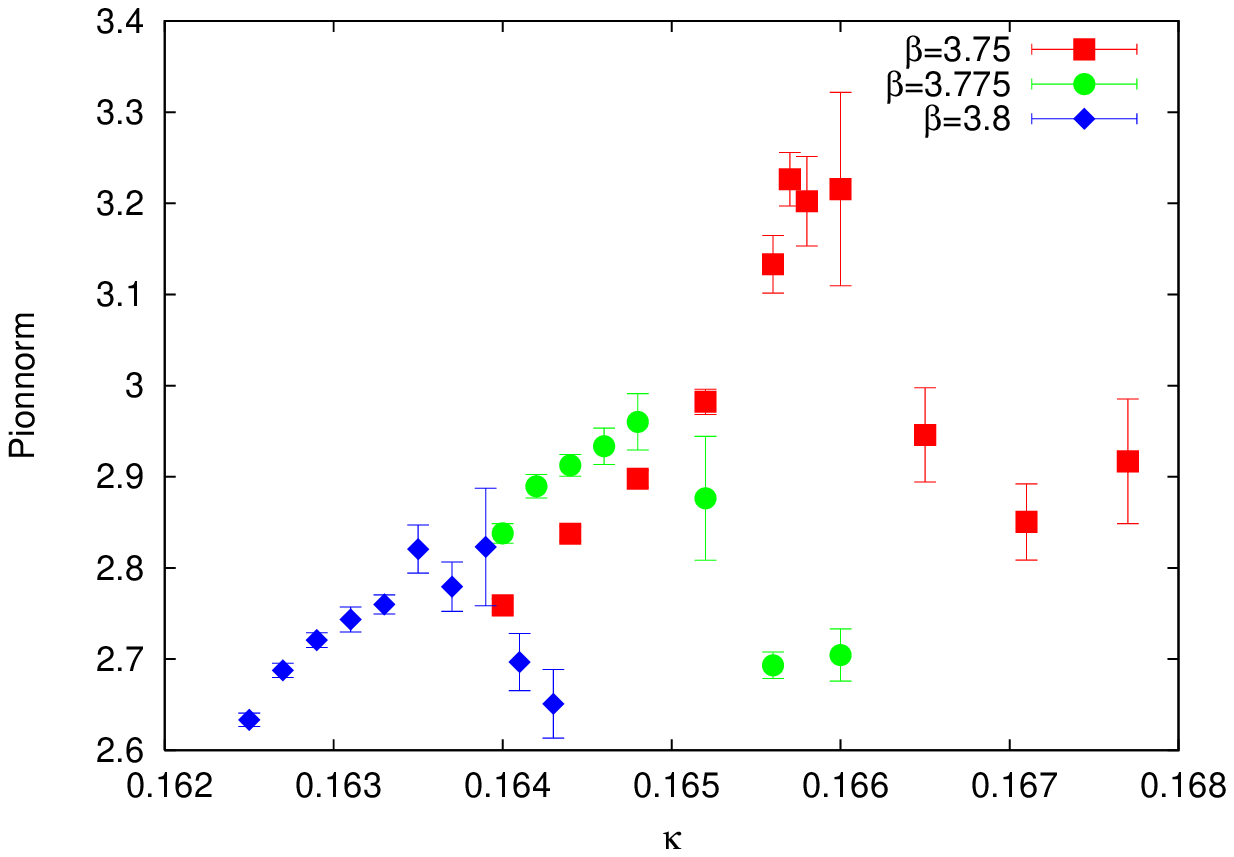} 
\end{minipage}
\begin{minipage}[t]{.45\textwidth}
\includegraphics[width=.9\textwidth]{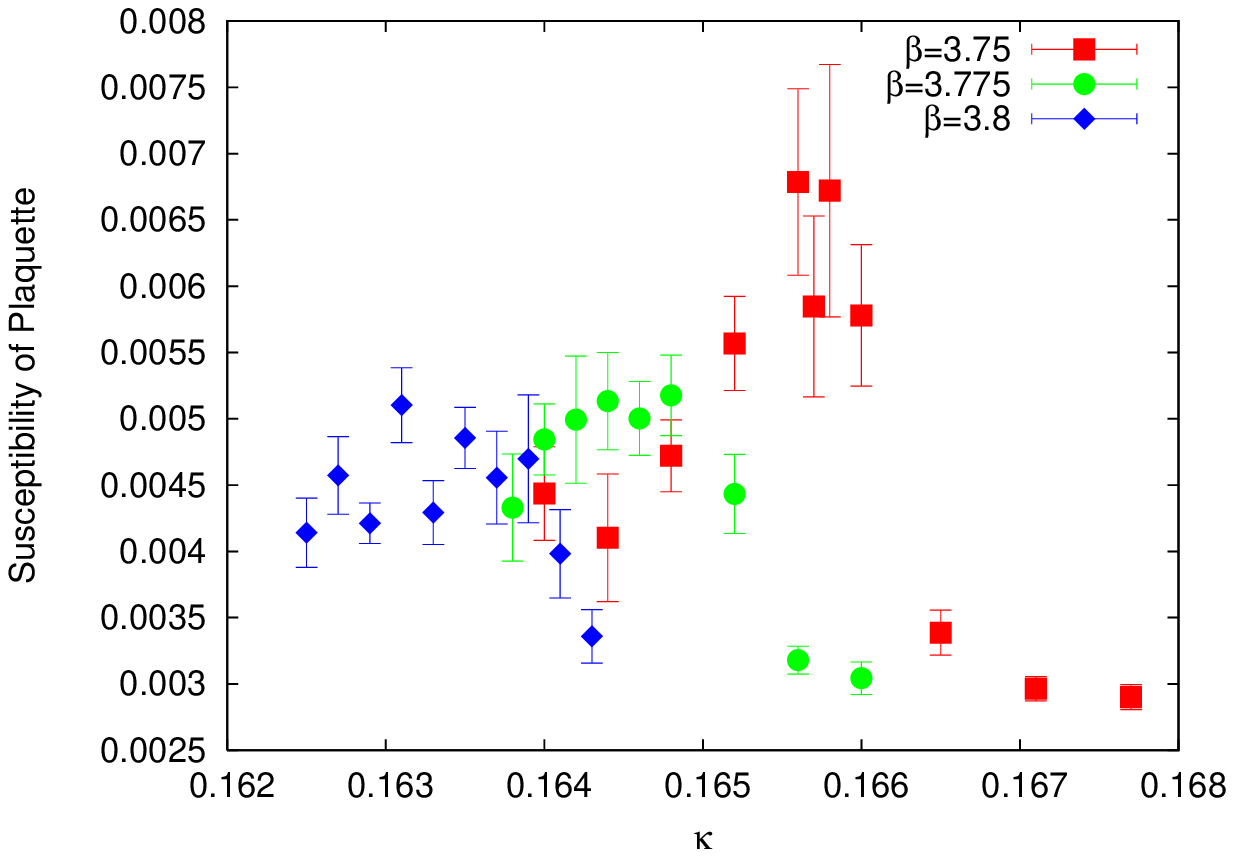} \\
\includegraphics[width=.9\textwidth]{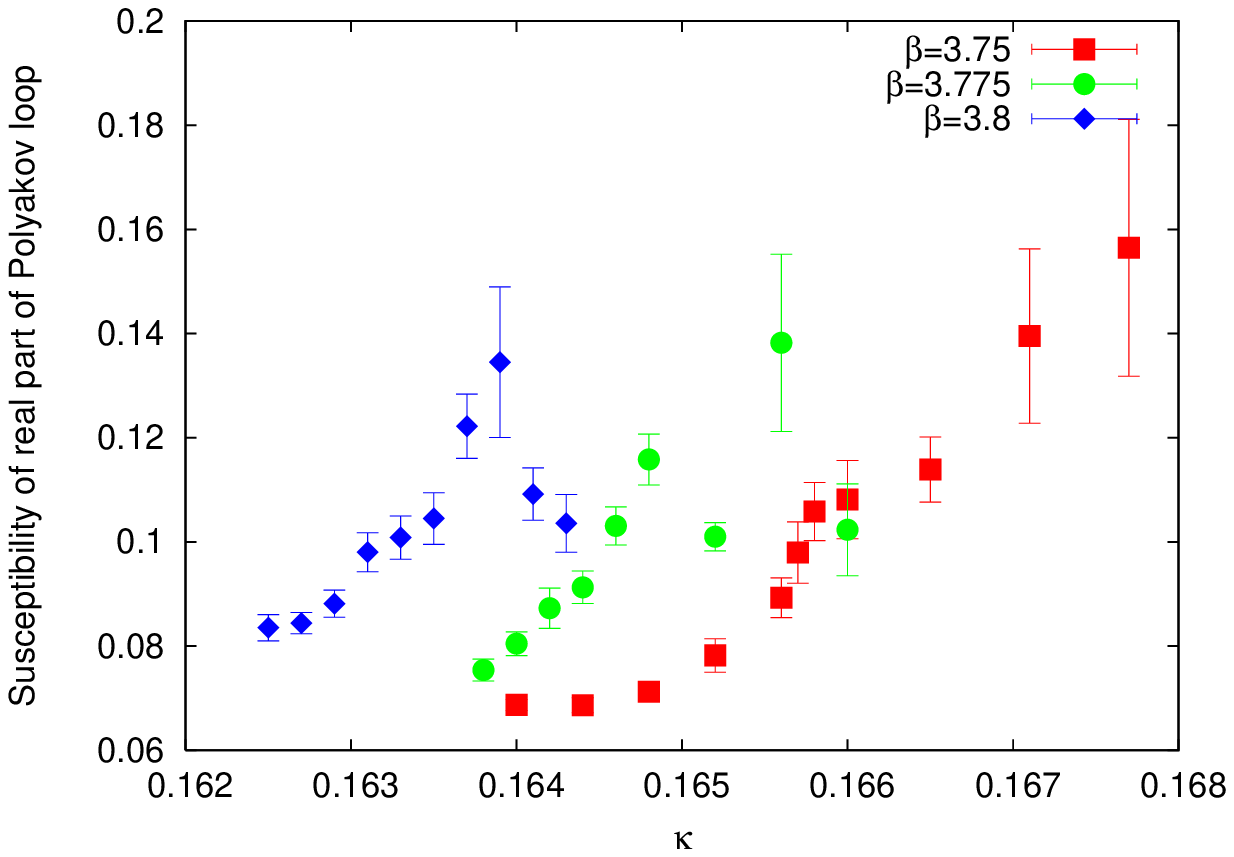}
\caption{Search for the thermal transition at $\mu=0.005$ and three values of 
$\beta$. \textbf{Left, from top to bottom:} Average plaquette, real part of the 
Polyakov loop and pion norm. \textbf{Right, from top to bottom:} Susceptibility of 
plaquette and real part of the Polyakov loop.\label{thermalfigs}}
\end{minipage}
\end{figure}

In Figure~\ref{thermalfigs} we present our data for the search for the
thermal transition line at $\mu=0.005$ for the various values of
$\beta$.  The critical $\kappa_T$ of the transition is decreasing with
growing $\beta$.  This is consistent with the fact that $T_c(m)$ is an
increasing function of quark mass, as found with staggered or
untwisted Wilson fermions (for a recent review, see
\cite{Philipsen:2007rj}). Correspondingly, the signal in the Polyakov
loop becomes more pronounced with increasing $\beta$, whereas the
signal in the plaquette becomes broader.  The values we extracted for
the thermal transition $\kappa_T(\mu,\beta)$ are collected in Figure~\ref{thermalline};
they are compared to the values for $\kappa_c(T=0,\beta)$,
cf.~\cite{Farchioni:2005ec,Boucaud:2007uk,Urbach:2007}.  As in the case of
untwisted Wilson fermions \cite{Ali Khan:2001ek}, the distance between
$\kappa_T$ and $\kappa_c$ grows with increasing $\beta$.

%\subsection{Conical Structure}
As stated above, due to equation~(\ref{eq:mass}), for fixed lattice
spacing, viz.~coupling $\beta$, there should be a line of equal physics
in the $(\kappa,\mu)$ plane. This leads to the expectation of a
conical thermal transition surface in the phase diagram as sketched in
Figure~\ref{creutzphase}.  In order to check this expectation we have
scanned for the transition both in $\kappa$ for fixed $\mu$, as well
as in $\mu$ for fixed $\kappa$.

\begin{figure}%[p]
\centering
\begin{minipage}[t]{.45\textwidth}
\includegraphics[width=.9\textwidth]{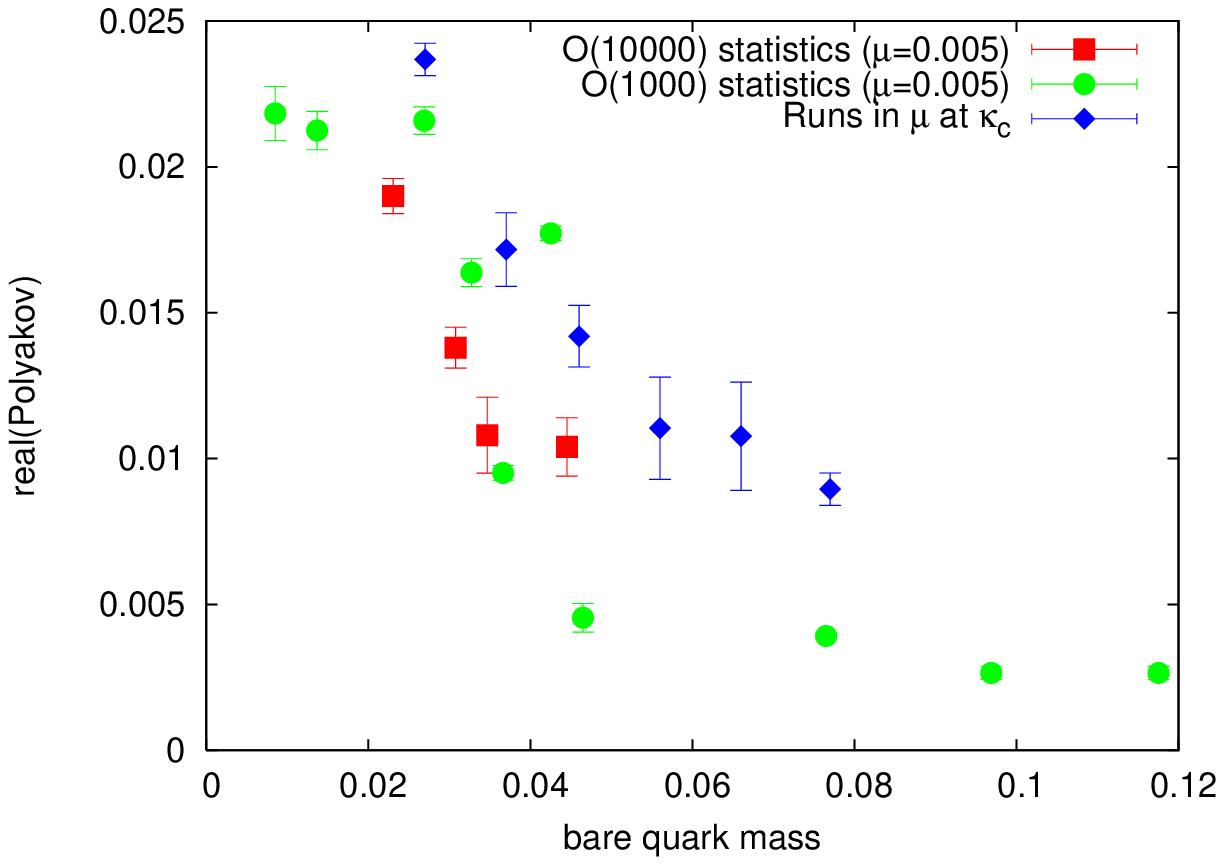} 
\caption{Comparison of runs for different $(\kappa,\mu)$ pairs at $\beta=3.9$ mapped to the 
bare quark mass.\label{comparison}}
\end{minipage}\hspace*{2mm}
\begin{minipage}[t]{.45\textwidth}
\includegraphics[width=.9\textwidth]{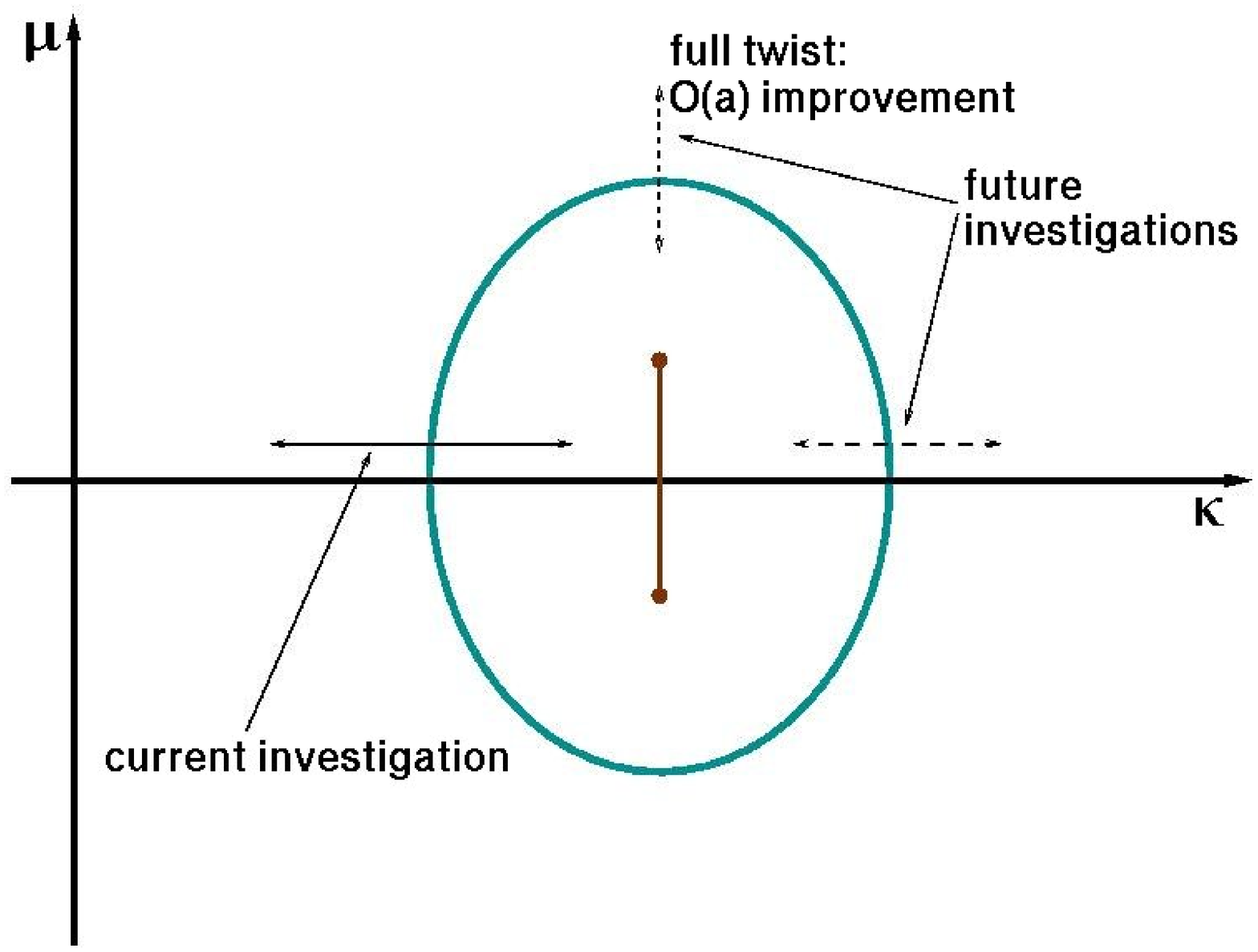}
\caption{Current and planned investigations in the context of the 
expected phase structure.\label{betaslice}}
\end{minipage}
\end{figure}

In Figure~\ref{comparison} we present the comparison of a scan in
$\mu$ that was performed at $\kappa=\kappa_c(T=0,\beta=3.9)$ with
$\mathcal{O}(4000)$ statistics with a scan in $\kappa$ at $\beta=3.9$
and $\mu=0.005$ with up to $\mathcal{O}(10000)$ HMC sweeps.  The
points are mapped to an effective bare quark mass using equation
(\ref{eq:mass}).  The similar behaviour provides preliminary evidence
for the existence of at least part of the conical structure of the
transition surface. Note that the data correspond to different
$\mu$-values. Since this varying of $\mu$, i.\,e. of the twist angle $\arctan\left(\mu/(m-m_c)\right)$, introduces $O(a)$ effects, the data
are distorted away from the unique curve by finite lattice spacing
effects.

In Figure~\ref{betaslice} we sketch an approximate slice of the phase diagram in
Figure~\ref{creutzphase} at a larger $\beta$. A line of equal critical mass, defined by (\ref{eq:mass}), for the
thermal transition is sketched.  The vertical line within the ellipse is the
$1^\mathit{st}$ order line containing $\kappa_c(\beta, T=0, \mu=0)$ as
predicted by $\chi PT$ for $T=0$ -- cf.~\cite{Munster:2004am}. Our
present and planned future investigations are illustrated as well.

\newpage

\vspace*{-1.4cm}
\section{Summary and Outlook}
For $\beta=3.4$ and $\mu=0$ we have found a very narrow range of
$\kappa$-values where we observe indications for a first order
phase transition similar to the bulk transition seen in the
zero-temperature case. %\cite{Sharpe:1998xm,Munster:2004am,Farchioni:2004us}.
There seems to be no Aoki phase in this range of the phase diagram.
The deconfinement transition occurs at considerably higher $\kappa$-values.

For larger $\beta$ we have identified the thermal transition line at
$\mu=0.005$, and provided some evidence for one side of the conical
phase boundary predicted in \cite{Creutz:2007fe}.

Our future work will consist of a continued investigation of the strong
coupling region at $\beta$-values substantially lower than presented here
as well as probing the details of the conical structure for larger $\beta$.
Finally, we would like to study the thermal transition at maximal twist
and to work towards physical quark masses and the continuum in this regime.

\vspace*{-0.2cm}
\acknowledgments
\vspace*{-0.1cm}
E.-M.\,I. is supported by DFG through the Forschergruppe 
Gitter-Hadronen-Ph\"anomenologie (FOR 465), A.\,S. by the Australian Research Council, and L.\,Z. thanks the BMBF Germany.
This work has been supported in part by the DFG
Sonderforschungsbereich/Transregio SFB/TR9-03. 
We thank the RM31 'Iniziativa Specifica' for time on the apeNEXT computer.

\end{document}